\documentstyle[aps,twocolumn,prl,epsf]{revtex}

\begin{document}

\draft

\twocolumn[\hsize\textwidth\columnwidth\hsize\csname @twocolumnfalse\endcsname

\title{Spinless Two-Band Model in Infinite Dimensions}

\author{Luis Craco}

\address{Max-Planck-Institut f\"ur Physik Komplexer Systeme, 
 N\"othnitzer Strasse 38, \\
D-01187 Dresden, Germany }

\date{\today}
\maketitle

\widetext
\begin{abstract}
\noindent

A spinless two-band model is studied in infinite dimension limit. 
Starting from the atomic limit, the formal exact solution of the 
model is obtained by means a perturbative treatment of the hopping 
and hybridisation terms. The model is solved in closed form in high 
dimensions assuming no local spin fluctuations. The non-Fermi liquid 
properties appearing in the metallic phase are analysed through the 
behaviour of the density of states and the self-energy near the 
Fermi level.
\end{abstract}

]

\narrowtext

In the last few years, deviations from the normal Fermi liquid (FL)
behaviour have been reported in a number of strongly correlated 
systems. These deviations were observed, for example, 
in the normal state of the 
high $T_c$ copper oxide superconductors~\cite{htc}, different
heavy-fermion compounds~\cite{heavy} as well as 
quasi-one-dimensional materials~\cite{qod}. This unexpected  
behaviour increased the interest in studying physical 
models that clearly present non-Fermi liquid properties.
Indeed, Consiglio and Gusm\~ao~\cite{Consiglio} have 
shown that the main features of the optical conductivity of the 
Kondo alloy $Y_{1-x}U_xPd_3$ are well taken into account by 
the simplified periodic Anderson model. Recently, the ac Hall
response for the normal state of cuprate superconductors~\cite{ML}
have been described by means of the simplified Hubbard 
(SH) model. 

On the other hand, several years ago  Si {\it et al.}~\cite{Si}
introduced a spinless two-band (SLTB) model to describe
the effect of interactions in a band insulating system~\cite{cmt}.
By solving the model numerically in the limit of 
infinite spatial dimensions $(d \rightarrow \infty)$ they showed 
that it exhibits NFL properties. In this letter we will present  
for the first time an analytical solution for this model in the 
same limit.  
  
The $d=\infty$ limit, introduced originally by Metzner and 
Vollhardt~\cite{MV}, has been shown to be a good starting point 
in the study of  several physical systems~\cite{Georges}. It is known 
that in this limit, local interactions give rise to a {\bf k}-independent 
self-energy and consequently the {\it starting} model can be mapped 
onto a single-site problem~\cite{GK}. Of the many different methods
applied to solve the single-site problem~\cite{Georges,LM2},
few of them allow us to obtain analytical results. Of these few methods 
only one allows us to obtain the single-site one-particle Green's 
function by means of a perturbative expansion in the local mean-field 
around the atomic limit~\cite{LM2}. This method, developed in the 
context of the Hubbard model, will be used here in 
order to solve the problem of local interactions between bands.

The SLTB model on a $d$-dimensional hypercubic lattice is described 
by the Hamiltonian

\begin{eqnarray}
\label{eq:model}
H & = & \sum_{\bf k} \epsilon_{\bf k} c_{\bf k}^\dagger c_{\bf k} 
+ \sum_{i} t  (d_{i}^\dagger c_{i} + h. c.) 
+ V \sum_{i} n_{i}^c n_{i}^d 
\nonumber \\ & + &
(\epsilon_{d}^0 - \mu) \sum_{i} n_{i}^d - \mu \sum_{i} n_{i}^c 
\end{eqnarray}
where $c_{i}^\dagger$ and $d_{i}^\dagger$ are the creation operators for the  
spinless electrons in the two different bands, 
$n_{i}^c = c_{i}^\dagger c_{i}$, 
$n_{i}^d = d_{i}^\dagger d_{i}$.  
$\epsilon_{\bf k}$ is the dispersion 
relation of the conduction ($c$) band and $\epsilon_{d}^0$ is the energy level
of the localised ($d$) electrons. According to Eq.~(\ref{eq:model}) these  two 
species of spinless electrons are coupled through the hybridisation $(t)$ 
and the interband correlation $(V)$ terms. 

The simplest solution of the Hamiltonian above is obtained in the 
noninteracting limit $(V=0)$. At half-filling the model presents a 
band-insulating behaviour: The bonding and the antibonding bands 
are split  by the hybridisation gap. In the $t=0$ limit, Eq.~(\ref{eq:model})
can be related to the SH model~\cite{SH}, which has been shown to 
possess an exact solution for $d=\infty$~\cite{BM}. In the paramagnetic 
phase the {\it conduction} electrons of the SH model exhibit a 
metal-insulator transition as a function of $V$. The 
dynamics of the localized electrons in this model is related to 
local spin fluctuations~\cite{LM2}. The corresponding single-site 
one-particle Green's function $(G_{ii}^d)$ is a non-trivial function of the 
dynamical mean field ${\cal A}_c$ of the conduction electrons~\cite{Janis}.
However, due to the spinless character of the SLTB model, where the spin flip, 
and double-occupied-to-empty site transitions~\cite{LM1} are excluded, 
one can assume that the spin fluctuation problem is absent 
in Eq.~(\ref{eq:model}). In this case, the dynamics of the $d$-electrons 
is governed by the hybridisation term. 

To solve the SLTB Hamiltonian we start by assuming that the local part 
of the unperturbed Hamiltonian is given by the last two terms of 
Eq.~(\ref{eq:model}). The solution 
of the {\it local} unperturbed Hamiltonian provides a basis of two decoupled 
spaces. At each site the interband correlation term in 
Eq.~(\ref{eq:model}) can be considered a perturbation of these 
decoupled spaces. For both $c$ and $d$ electrons this problem is   
equivalent to the atomic problem of the Hubbard model~\cite{LM1},  
and the corresponding zero-order Green's functions are given by

\begin{equation}
\label{eq:g0}
{\cal G}_{0}^\alpha (i\omega_n) = 
\frac{1-\langle n_\gamma \rangle_0}{i\omega_n +\epsilon_\alpha}
+ \frac{\langle n_\gamma \rangle_0}{i\omega_n +\epsilon_\alpha - V}
\;,
\end{equation}
where $\alpha$ and $\gamma$ stand for $c$ or $d$ and 
$\alpha \ne \gamma$, $\epsilon_c \equiv \mu$ and 
$\epsilon_d \equiv \mu -\epsilon_{d}^0$. 

Next we consider the $t=0$ limit of Eq.~(\ref{eq:model}). In this limit 
the hopping term provides dynamics only for the $c$-electrons. 
The solution to this particular problem can be obtained by 
means of a tight-binding treatment around the atomic limit of the 
conduction electrons~\cite{LM1}. The correspondent single-particle 
Green's function is given by 

\begin{equation}
\label{eq:g0ck}
{\bar g}_{\bf k}^c (i\omega_n) = \frac{1}{ [{\cal G}_{\bf k}^c (i\omega_n) ]^{-1} 
-\epsilon_{\bf k}} \;, 
\end{equation}
where ${\cal G}_{\bf k}^c (i\omega_n)$ is the so called irreducible
one-particle Green's function. ${\cal G}_{\bf k}^c (i\omega_n)$ is
irreducible in the sense that  it can not be divided in two pieces by 
cutting a single hopping line.  

Finally, the formal exact solution of the 
complete problem is obtained by performing a perturbative treatment 
on the hybridisation term~\cite{Consiglio}. The perturbation series for 
both $c$ and $d$ one-particle Green's functions can be written 
as~\cite{matrix} 

\begin{equation}
\label{eq:gck}
G_{\bf k}^c (i\omega_n) = {\bar g}_{\bf k}^c (i\omega_n) + 
{\bar g}_{\bf k}^c (i\omega_n)  t {\cal G}_{\bf k}^d (i\omega_n) t 
G_{\bf k}^c (\i\omega_n) 
\end{equation}
and
\begin{equation}
\label{eq:gdk}
G_{\bf k}^d  (i\omega_n) = {\cal G}_{\bf k}^d (i\omega_n) + 
{\cal G}_{\bf k}^d (i\omega_n)  t G_{\bf k}^c (i\omega_n) t 
{\cal G}_{\bf k}^d (\i\omega_n) \;.
\end{equation}

In these two equations, ${\cal G}_{\bf k}^d (i\omega_n)$ is the 
irreducible one-particle Green's function of the $d$-electrons. 
As in the case of the Anderson model~\cite{Consiglio}, 
${\cal G}_{\bf k}^d (i\omega_n)$ is {\it irreducible} in the sense 
that it can not be divided in two parts by the process of cutting 
a $t^2 {\bar g}_{\bf k}^c (i\omega_n)$ line. Note that in our formalism, 
it is not necessary to define the non-diagonal self-energies~\cite{Georges}, 
whose explicit form are normally unknown quantities. 

In the limit of infinite spatial dimensions, only the site-diagonal 
part of the irreducible propagator survives~\cite{LM2}. In this case 
the ${\bf k}$-sum of the Eqs.~(\ref{eq:gck}) and~(\ref{eq:gdk}) can be 
easily performed and it is straightforward to show that 

\begin{equation}
\label{eq:gcii}
G_{ii}^c  (i\omega_n) =\frac{1}{N}\sum_{\bf k}  \frac{1}{ 
[ {\cal G}^c (i\omega_n) ]^{-1} 
-\epsilon_{\bf k} - t^2 {\cal G}^d (i\omega_n) }
\end{equation}
and
\begin{equation}
\label{eq:gdii}
G_{ii}^d  (i\omega_n) = {\cal G}^d (i\omega_n) [ 1 +  
{\cal G}^d (i\omega_n)  t^2 G_{ii}^c (i\omega_n) ] \;.
\end{equation}
 
The presence of an interband interaction $V$  in 
Eq.~(\ref{eq:model}) causes the appearance  
of two irreducible propagators (one for each band) in the 
Eqs.~(\ref{eq:gck}) and~(\ref{eq:gdk}). In the high-dimensional limit,  
one can write the irreducible propagators in terms of the single-site 
one-particle Green's function and the dynamical mean-field through the 
relation~\cite{LM2}

\begin{equation}
\label{eq:aa}
\frac{1}{ {\cal G}^\alpha (i\omega_n) } = 
\frac{1}{ G_{ii}^\alpha (i\omega_n) } + {\cal A}_\alpha (i\omega_n) \;,
\end{equation}
where the index $\alpha$ represents both $c$ and $d$ 
electrons~\cite{general}. 

In order to have the complete set of equations for the relevant  
Green's functions as well as the dynamical mean fields, an explicit 
solution of the single-site problem for each band must be obtained.
The perturbation method around the atomic limit provides a direct  
way of solving the single-site problem by means a perturbative 
expansion in the local mean-field~\cite{LM2}. Starting from the 
unperturbed local Green's functions, Eq.~(\ref{eq:g0}), 
two perturbation treatments in local mean fields must be performed. 
Since we have assumed that the local unperturbed basis 
is a product of two decoupled spaces, only contractions between 
one-band and one-site electron operators are non-vanishing. 
This means that in each band the local contractions give always the 
same factor $g_\alpha (i\omega_n)=1/(i\omega_n +\epsilon_\alpha)$, 
with $\alpha=c$ or $d$, which describe the excitation energy when 
only one level  is occupied.  When both single-site levels are 
occupied the excitation energy is shifted by $V$, as one can see 
from Eq.~(\ref{eq:g0}). In this case the perturbative 
expansion in the local mean field can be summed exactly, 
which provides a renormalisation in 
$g_\alpha $ of the form $g_\alpha (i\omega_n) = 
1/(i\omega_n +\epsilon_\alpha - {\cal A}_\alpha (i\omega_n) )$. 
Following this procedure one can show that the single-site 
one-particle Green's functions for both electrons is given by 
 
\begin{equation}
\label{eq:ga}
G_{ii}^\alpha  (i\omega_n) = 
\frac{1-\langle n_\gamma \rangle}{i\omega_n +\epsilon_\alpha 
- {\cal A}_\alpha  (i\omega_n) }
+ \frac{\langle n_\gamma \rangle}{i\omega_n +\epsilon_\alpha - V
- {\cal A}_\alpha  (i\omega_n) } \;.
\end{equation}

Eqs.~(\ref{eq:gcii})-(\ref{eq:ga}) constitute a closed 
system of equations 
for the SLTB model in infinite dimensions. From  Eqs.~(\ref{eq:gdii}) 
and~(\ref{eq:aa}) one can easily see that 

\begin{equation}
\label{eq:adfor}
{\cal A}_d (i\omega_n)=\frac{t^2 {\cal G}^c (i\omega_n)}{
1-({\cal A}_c (i\omega_n) - t^2 {\cal G}^d (i\omega_n)) 
{\cal G}^c (i\omega_n)} \;.
\end{equation}
In this equation $t^2  {\cal G}^c (i\omega_n)$ describes the loop 
contribution of the hybridisation term, while 
$({\cal A}_c (i\omega_n) - t^2  {\cal G}^d (i\omega_n))$ represents
all simple loops coming from the hopping process 
($t^2  {\cal G}^d (i\omega_n)$ avoid the over-counting of the   
hybridisation loops). Furthermore,  
since ${\cal A}_d (i\omega_n) = 0 $ for $t=0$, 
$G_{ii}^d (i\omega_n) = {\cal G}_0^d (i\omega_n)$ in this particular 
limit (see  Eq.~(\ref{eq:ga})). As we have mentioned before, the purely 
atomic character of $d$-electrons when $t=0$ is one of the most 
important differences between the present solution of the SLTB model 
and the solution of the SH model~\cite{Janis}. 

\begin{figure}[t]
\epsfxsize=3.2in
\epsffile{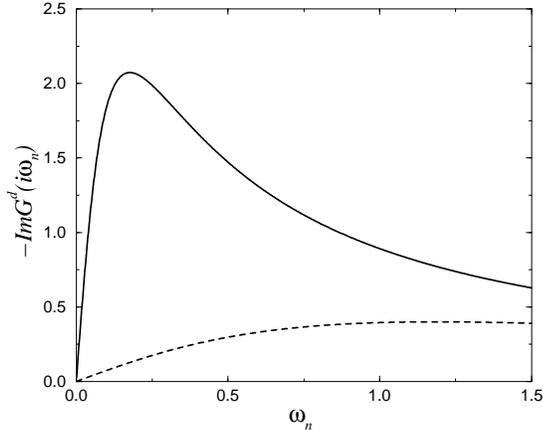}
\caption{$G_{ii}^d  (i\omega_n)$ vs Matsubara frequency $\omega_n$ 
for $t=0.4$ and two values of $V$: $V=-0.1$ (continuous) and 
$V=-2.5$ (dashed).} 
\label{fig1}
\end{figure}

The self-consistent solution of Eqs.~(\ref{eq:gcii})-(\ref{eq:ga})
can be performed numerically for Matsubara as well as for 
real \mbox{$(i\omega_n \rightarrow \omega + i0^+)$} frequencies. 
In what follows, we will consider the case of a semicircular uncorrelated 
density of states for the conduction band at half-filling, where 
$\epsilon_{d}^0=0$ and $\mu=V/2$. In Fig.~\ref{fig1} we plot 
the imaginary part of the $d$-electron Green's function as a 
function of the Matsubara frequencies for $t=0.4$ and two values of 
$V$. For these parameters one can see that our analytical results agree 
well with those obtained numerically by Si {\it et al.}~\cite{Si}. 

It is known that in the limit of small $t$ and large values of $V$, 
the hybridisation is irrelevant in the renormalisation group 
sense~\cite{Si}. This explain the good agreement between our results 
for $V=-2.5$ (dashed line of Fig.~\ref{fig1}) and those obtained by 
Si {\it et al}. Note that in this limit, the solution of the SLTB model 
becomes similar to that of the spinless Falicov-Kimball model~\cite{Si-FK}
in the non-Fermi liquid regime.

\begin{figure}[h]
\epsfxsize=3.2in
\epsffile{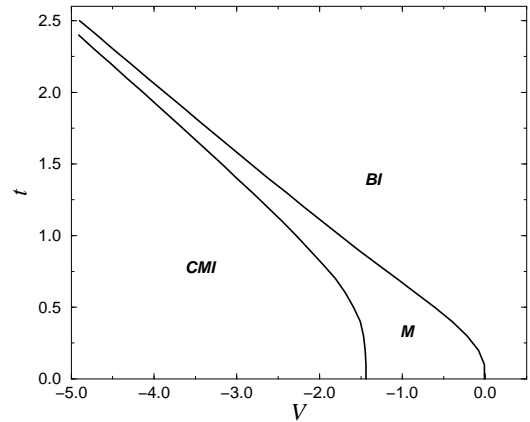}
\caption{The phase diagram of the SLTB model.
In this figure $BI$, $M$ and $CMI$ are the band insulating, 
metallic, and charge Mott insulating phases, respectively.} 
\label{fig2}
\end{figure}

Let us now turn our attention to Fig.~\ref{fig2}, 
where we show the phase diagram of the SLTB model. 
As was pointed out in Ref.~\cite{Si}, this model exhibits three different 
phases: the band insulating (BI), the metallic (M) and the charge 
Mott insulating (CMI) phase. For $t=0$ the system goes from a metal 
into a charge Mott insulating phase as in the SH model. 
For largest values of $t$ and $V$, the separation between the metallic
and the insulating phases increases linearly, around the line 
$V=-2t$.

\begin{figure}[hbt]
\epsfxsize=3.2in
\epsffile{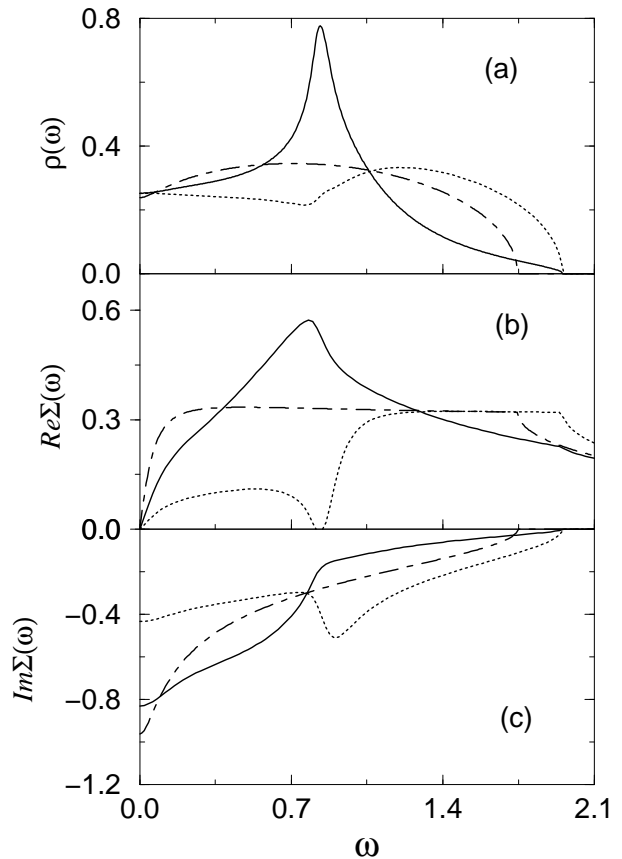}
\caption{Single-particle properties of the $d$-electrons (solid line), 
the $c$-electrons (dotted line)  for $V=-2t=-1.2$. For comparison we 
also include the spectral properties of the conduction electrons of the 
SH model (dot-dashed line). The DOS is shown in Fig. 3-a.
The real and imaginary parts of the self-energies are plotted in 
Fig.~3-b and~3-c, respectively.} 
\label{fig3}
\end{figure}

The different dynamics of the two types of electrons 
together with the spinless nature of the model 
Hamiltonian (Eq.~(\ref{eq:model})) make the metallic phase 
a non-Fermi liquid one. In Ref.~\cite{Si} the non-Fermi liquid properties 
of the SLTB model were studied analysing the 
behaviour of the single-site one-particle Green's functions
of the $c$ and $d$ electrons at Matsubara frequencies. 
Let us now proceed to analyse these properties in real space,  
on the line $V=-2t$. At this line, the antibonding, empty 
and doubly occupied states are degenerated and the metallic 
phase persists even for large values of $t$ and $V$. It is worth 
noticing that for all values of $V=-2t$ the density 
of states (DOS) of both electrons coincide at Fermi level. This can be
seen in Fig.~\ref{fig3}-a, where the DOS of $c$ and $d$ electrons are 
shown for $V=-2t=-1.2$.  

As can be seen in Fig.~\ref{fig3}-a the DOS of the $d$-electrons presents
a pronounced satellite peak reminiscent of the atomic $(t=0)$ limit. 
Moreover, the DOS of both electrons at the Fermi level is not 
pinned at the unperturbed one~\cite{comm}, as one would expect from 
a non-Fermi liquid solution. In Fig.~\ref{fig3}-b and Fig.~\ref{fig3}-c 
we show the real and imaginary parts of one-particle self-energy 
of $c$ and $d$ electrons. These self-energies can be defined by means 
of Eqs.~(\ref{eq:gck}) and~(\ref{eq:gdk}) as $\Sigma_\alpha(\omega) 
\equiv  \omega - \left[ {\cal G}^\alpha (\omega) \right]^{-1}$. 
In Fig.~\ref{fig3}-b one can see that the real parts of $\Sigma_c (\omega)$  
and $\Sigma_d (\omega)$ have a positive slope at $\omega \rightarrow 0$.
It is known that this slope must be negative for a Fermi liquid system.
Hence, our results confirms the breakdown of the quasiparticle concept, 
which can be also seen in Fig.~\ref{fig3}-c. 
From this figure it is clear 
that the imaginary part of $c$ and $d$ self-energies do not approach 
the quadratic Fermi liquid form close to the Fermi level. If one compares  
the self-energy of the SH model with the one of the $c$-electrons  
in the low energy region, one can see that the latter is strongly modified 
by the hybridisation. The slope of the $Re \Sigma_c (\omega)$ 
as well as the values of the  $Im \Sigma_c (\omega)$ are reduced 
at $\omega \rightarrow 0$. Such modifications reflect the localisation of 
the $c$-electrons due to the hybridisation term. The similarities between 
the self-energy of $d$-electrons and the SH model one 
indicate that the hybridisation term for the parameters of Fig.~\ref{fig3} 
is able to transfer almost completely the dynamics of the conduction 
electrons to the localised ones. 

Summarising, we present here for the first time an analytical solution
for the SLTB model. Assuming the absence of the spin fluctuation problem, 
we were able to derive a closed set of equations at the limit of high dimensions.
These equations were then studied at the real and Matsubara frequencies. 
In the latter case our results agree with those obtained numerically 
by Si {\it et al}~\cite{Si}. This result opens several possibilities to extent
this work, for example, the case of the spinful 
version of this two band model. Closely related to this point  
is the possibility of recovering the FL properties of a spinful system 
from the point of view of a cumulant expansion. Finally, one can also 
consider the SLTB model as a one band model~\cite{SH}.  
From this point of view it could be interesting to study how are the 
dynamic properties of the conduction electrons modified by the presence 
of such magnetic field.

The author wishes to acknowledge Prof. P. Fulde 
for advice and hospitality at the MPIPKS as well as 
M. A. Gusm\~ao, M. S. Laad, S. Blawid and J. L. Vega 
for useful discussions.


\begin{references}
\bibitem{cmt} The problem is related with the semi-metallic
nature of Kondo insulators~\cite{KI}, more precisely, whether 
interactions are able to provide a metallic phase in hybridisation 
gap alloys.
\bibitem{matrix} Eqs.~(\ref{eq:gck}) and~(\ref{eq:gdk}) are the diagonal 
($c$ and $d$ Green's functions, respectively) components of the 
matrix Green's function for the SLTB model.
\bibitem{comm} In the case of semicircular uncorrelated conduction band, 
the DOS at Fermi level is $\rho_0 (E_f)=0.45$.
\end{references}
\end{document}